\documentclass[aps,prl,reprint,amsmath,amssymb]{revtex4-1}

\usepackage{graphicx}
\usepackage{dcolumn}
\usepackage{bm}
\usepackage{physics}
\usepackage[dvipsnames]{xcolor}

\usepackage[mathlines]{lineno}

\begin{document}

\preprint{AAPM/123-QED}

\title[Curvature-mediated Programming of Liquid Crystal Microflows]{Curvature-mediated Programming of Liquid Crystal Microflows}

\author{K. Fedorowicz$^1$}
\author{R. Prosser$^1$}%
\author{A. Sengupta$^2$}
 \email{anupam.sengupta@uni.lu}
\affiliation{%
$^1$Department of Mechanical, Aerospace \& Civil Engineering, The University of Manchester, Manchester M13 9PL, UK.\\
$^2$Physics of Living Matter, Department of Physics and Materials Science, University of Luxembourg, 162 A, Avenue de la Fa\"{i}encerie, L-1511 Luxembourg City, Luxembourg
}%

\date{\today}

\begin{abstract}

Using experiments and numerical simulations, we demonstrate that the curvature of microscale conduits allow programming of liquid crystal (LC) flows. Focusing on a nematic LC flowing through U- and L-shaped channels of rectangular cross-section, our results reveal that the curvature of flow path can trigger transverse flow-induced director gradients. The emergent director field feeds back into the flow field, ultimately leading to LC flows controlled by the channel curvature. This curvature-mediated flow control, identified by polarizing optical microscopy and supported by the nematofluidic solutions, offers concepts in LC microfluidic valves, wherein the throughput distribution is determined by the Ericksen number and variations in local curvature. Finally, this work highlights the role of deformation history on flow-induced director alignments, where the viscous and elastic effects comparable in strength.

\end{abstract}

\maketitle

\section{Introduction}


Microfluidics, the science and technology of flow manipulation at microscales \cite{bruus2008theoretical, Whitesides2006}, touches every sphere of our lives today \cite{pfohl2003,Kunti2019,Paratore2021}. Microfluidic devices combine several laboratory processes into a single low-footprint instrument; their compactness, reduced analysis cost and lower testing times make them a potential competitor for current large-footprint devices \cite{Dressler2014,kwang2020,bebe2002,Priya2021}. Although traditionally, microfluidics has been based on isotropic fluids, either Newtonian or complex fluids, recent advances in the field of liquid crystal microfluidics \cite{sengupta2014} have demonstrated unique potential of LCs in creating tunable fluid flows and fluidic properties at microscales \cite{Sengupta2013,Sengupta2013_cargo,Ravnik2013,Kos2020}. 

Nematic liquid crystals (NLCs), mesogens comprising rod- or disc-like molecules \cite{stewart2004static}, combine liquid fluidity with crystalline properties, thus offering distinct attributes which arise due to the coupling between the molecular alignment (referred to as the director field $\vb{n}$ \cite{physicsOfLiquidCrystals}), hydrodynamics and confinement effects \cite{sengupta2014, xu2021}. LCs form systems with a non-zero orientational order, and the resultant material properties (hydrodynamic, optical, electrical etc.) depend on the mean local orientation. Controlling the emergent flow-behaviour via the interplay of elastic, viscous, and surface forces enable unique dynamical properties, offering mechanisms to manipulate LC flows\cite{Khoshbin2021, Woltman2007, Fedorowicz2023_liqCryst}. The director orientations on the confining surfaces and within the bulk control the properties of NLC flows \cite{Pieranski1974,sengupta2014}. Flow-dependent LC viscosity has already been exploited in non-intrusive flow control devices \cite{Na2010,Sengupta2013-2}, as has their orientation-dependent birefringence, which provides the functionality for colour filters \cite{Cuennet2013}, optical \cite{Sengupta2012, Eichler2019} and biochemical \cite{liu2012} sensors. The behaviour of the NLC flows in rectangular channels \cite{Sengupta2013, copar2020} emerges due to the competition between elastic and viscous effects, captured by the dimensionless Ericksen number, $Er$ \cite{physicsOfLiquidCrystals}. 

\begin{figure}
\includegraphics[width=1\linewidth]{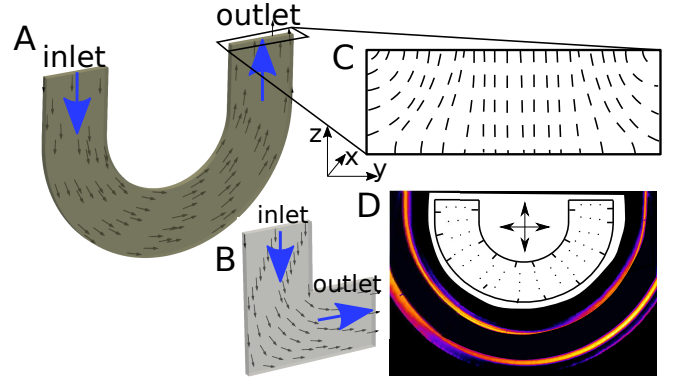}
\caption{\label{paper_fig1_3} Schematic description of microfluidic channels considered in this paper (A) U-channel; (B) L-channel; (C) static director field at the outlet of the U-channel; (D) light intensity under crossed polarizers and director arrangement in the centre plane (inset); dots denote the director orientation perpendicular to crossed polarizers (arrows). }
\end{figure}


Historically, NLC studies were motivated by the development of liquid display devices, where \textit{flat} channels (width$\gg$height) simplified the problem to two dimensions \cite{Denniston2001, denniston2001b,softmatterbook,Quintans2006, cruz2013,Rey2002,fedorowicz2022_contr}, however with limited relevance to \textit{real} microfluidic settings, where the  channel height and width are comparable\cite{Sengupta2013, Sengupta_thesis, Mondal2018}. While these recent efforts have successfully attained tunable LC flows by combining appropriate surface anchoring, confinements, and hydrodynamic forcing, the role of channel curvature has remained unexplored. In comparison, flow of isotropic fluids through curvilinear microchannels have received extensive attention for many decades, with distinct curvature-induced fluid dynamic outcomes \cite{Deam1928,Dicarlo2009,Steinberg2021}. More generally, our current understanding of LC flows through complex microfluidic geometries is limited to a handful of studies, for instance within rectangular channels with obstacles \cite{SenguptaMicropillar}, and at topological flow junctions \cite{giomi2017}. Numerical investigations into three-dimensional capillary bends \cite{Fedorowicz2022_pof} and junctions \cite{kos2017} have been reported, yet they currently lack experimental validation.


In this letter, we use a combination of LC microfluidic experiments and numerical methods to demonstrate novel flow control mechanisms in curvilinear microfluidic channels of varying complexity (fig. \ref{paper_fig1_3}A, B). The quasi-Newtonian behaviour observed at low Ericksen number regime $Er<1$ is altered as the Ericksen number exceed $Er>1$, along with the emergence of a flow-induced director field gradient in the transverse channel direction. The transverse gradient, introduced by the channel curvature, engenders a flow-partitioning effect due to the backflow mechanism \cite{Sengupta2013}, which is amplified by the viscous anisotropy of liquid crystals. Finally, we show that the deformation history affects the steady-state director and flow distribution at intermediate values of $Er \approx 15$. Taken together, the curvature- and history-mediated modulation of NLC flow properties can be harnessed for potential concepts and design of LC-based microfluidic throttles and valves. 



\begin{figure*}
\includegraphics[width=1\textwidth]{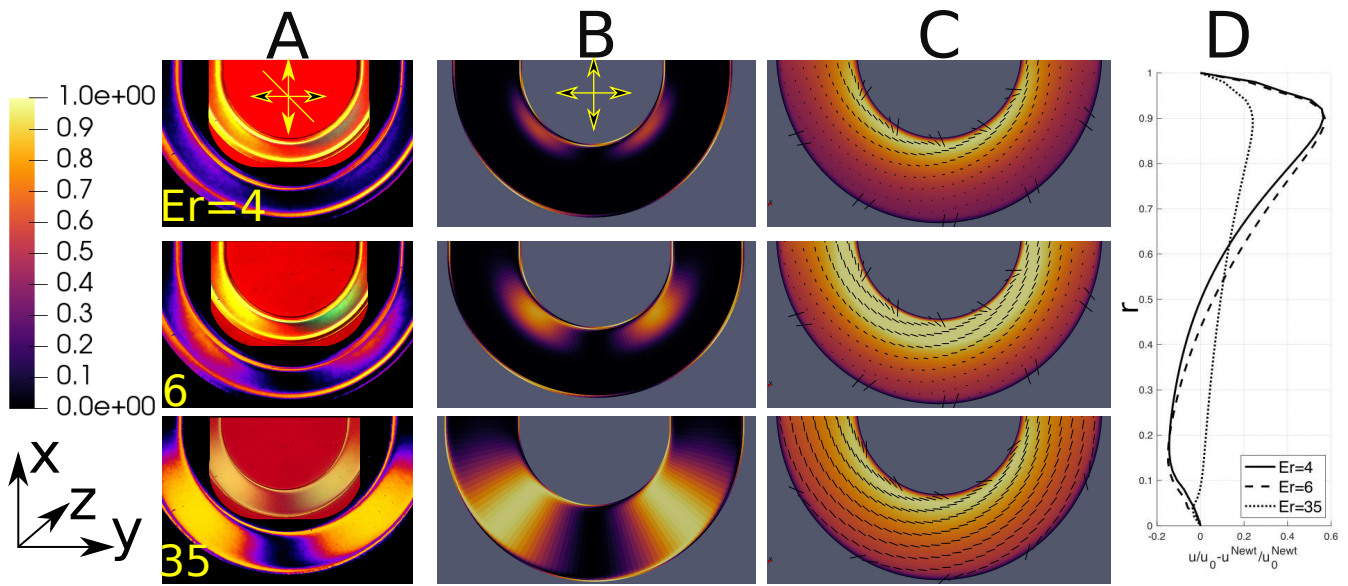}
\caption{\label{combined_u_bend2} Director alignment in a flow through U-bend. (A) Experimental micrographs showing transmitted light intensity under crossed polarizers. Inset: Polarized micrograph with a $\lambda$-retardation plate; the yellow and the blue-green signals respectively denote director alignment parallel and perpendicular to the retardation plate. (B) Numerical prediction of intensity under crossed polarizers (calculated as $4n_x^2n_y^2$). (C) Contours of flow magnitude scaled by the maximum flow speed and glyphs of the director field. (D) Effect of nematic liquid crystals on the amplification of flow asymmetry (in the symmetry of the U-bend) with respect to a Newtonian fluid. $r$ denotes the normalised distance from the outer bend wall.}
\end{figure*}

\section{Methodology}
\subsection{Simulations}
The flow of an NLC is assumed to be solenoidal and governed by the momentum equation, which reads
\begin{equation}
\label{momentumEquation}
\rho ( \partial_t + u_k \partial_k) u_i = -p_{,i} + \mu u_{i,jj} + \tau_{ji,j}  ,
\end{equation}
where $p_{,i}$ is the pressure gradient, $\mu$ is the Newtonian viscosity and $\tau_{ji}$ is the viscoelastic stress \cite{beris1994}
\begin{eqnarray}
\tau_{ij} &=&  H_{ik} Q_{kj} - Q_{ik} H_{kj}  + 2 \xi( Q_{ij}+\delta_{ij} ) H_{kl}Q_{kl} \nonumber\\
&-& \xi \big[ ( Q_{ik}+\delta_{ik} )H_{kj} + H_{ik}( Q_{kj}+\delta_{kj} ) \big]  \nonumber\\
&-& K Q_{kl,i} Q_{kl,j} .
\end{eqnarray}
$K$ is the elastic constant and $Q_{ij}$ is the order parameter tensor, which evolves according to \cite{beris1994}
\begin{equation}
    \partial_t^{\square} Q_{ij} = \Gamma H_{ij} - 2 \xi ( Q_{ij} + \delta_{ij} ) Q_{kl} u_{k,l}.
\end{equation}
$\partial_t^{\square}$ represents the Gordon-Schowalter derivative with slip parameter $\xi$, and $\Gamma$ is the rotational diffusivity. The molecular field $H_{ij}$ is defined in terms of the variational derivative of the Helmholtz free energy $F$\cite{physicsOfLiquidCrystals}:
\begin{equation}
F = \frac{K}{2} Q^2_{ij,k} + \frac{a}{2} Q_{ij}Q_{ij} - \frac{b}{3} Q_{ij}Q_{jk}Q_{ki} + \frac{c}{4} (Q_{ij}Q_{ij})^2,
\end{equation}
where $a$, $b$ and $c$ are material parameters. The relative importance of viscous and elastic effects acting on the microstructure is quantified by the Ericksen number $Er=u_0 h/(\Gamma K)$, where $u_0$ and $h$ are velocity and length scales characterising the flow.

The governing equations describing liquid crystal behaviour were solved with the OpenFOAM solver \textit{rheoFoamLC}\cite{Fedorowicz2022_pof, github}, initialised with an isotropic microstructure and static flow conditions ($\vb{Q} = \vb{0}, \ \vb{u}=0$). No-slip conditions for the velocity and strong homeotropic anchoring were imposed on all walls. The fluid enters the channel with a uniform velocity at the inlet. The following values of material parameters were used for all simulations; $\mu=0.2 \text{Pa}\cdot \text{s}$, $K=40 \text{pN}$, $a=-2 \text{J}/\text{m}^3$, $b=40 \text{J}/\text{m}^3$, $c=40 \text{J}/\text{m}^3$, $\xi=1$, $\Gamma=7.29 (\text{Pa}\cdot \text{s})^{-1}$. These parameters ensure that the Deborah number ($=\frac{u_o}{\Gamma b h}$) is much smaller than unity $De<<1$, so defect properties were weakly affected by the flow \cite{kos2017}.

\subsection{Experiments}
Numerical simulations are complemented with microfluidic experiments, carried out within custom-built U- and L-shaped microchannels having rectangular cross-sections. The microchannels were assembled using polydimethylsiloxane (PDMS) molds and glass plates, bonded together after a brief plasma exposure \cite{sengupta2012func}. The assembled channels were first treated with an aqueous solution of the silane DMOAP, that induced homeotropic surface anchoring condition on the inner walls. For the current study, we report experiments with a channel height of 15 $\mu$m and width of 100 $\mu$m. A single component thermotropic LC (5CB), which is nematic at room temperature, was flowed into the microchannels through cylindrical tubes embedded into holes punched through the PDMS. A precise and controlled LC flow inside the microfluidic channels was achieved by pushing a gas-tight syringe pump using a microfluidic gear pump (with a flow resolution of nl/h). For anchoring characterisation, the functionalized channels were filled with 5CB in the isotropic phase, which was first allowed to cool down to a nematic phase. No perceptible swelling of the PDMS by the 5CB was observed during the course of experiments, in agreement with the observed dependence of the swelling ratio on the substance polarity \cite{sengupta2012func, sengupta2014constraints}. We characterize the surface- and buld director fields using a combination of polarized optical imaging (POM), with and without the $\lambda$-retardation plates. The transmitted light depends on the orientation of the director field relative to the polarizer axes, following Refs. \cite{sengupta2012func,Sengupta_thesis}.


\section{ Results }

\subsection{Analytical predictions }
\label{analtyical_prediction}
For a steady-state, fully-developed channel flow (with the cross-section shown in fig. \ref{paper_fig1_3}C), the non-dimensionalised angular momentum balance equation for a uniaxial liquid crystal in a very wide channel ($y>>z$) can be simplified to a non-linear Poisson equation
\begin{equation}
\label{director_balance}
(\partial_{yy}+\partial_{zz}) \theta + u_{x,z} Er \sin^2 \theta = 0,
\end{equation}
where $\theta$ is the angle between the director and the flow direction (eq. (\ref{director_balance}) is derived in the supplementary material). When $Er<<1$, the homeotropic anchoring imposed at the walls forces the director to align in the $y-z$ plane, normal to the flow direction. Viscous effects become significant when $Er=O(1)$, and their action is manifested by reorientation of the director field towards the flow direction $\theta \to 0$. Eq. (\ref{director_balance}) suggests that the effect of Ericksen number is amplified by the local velocity gradient; thus, the flow alignment is promoted in regions of elevated shear rate.





\begin{figure*}
\includegraphics[width=1\textwidth]{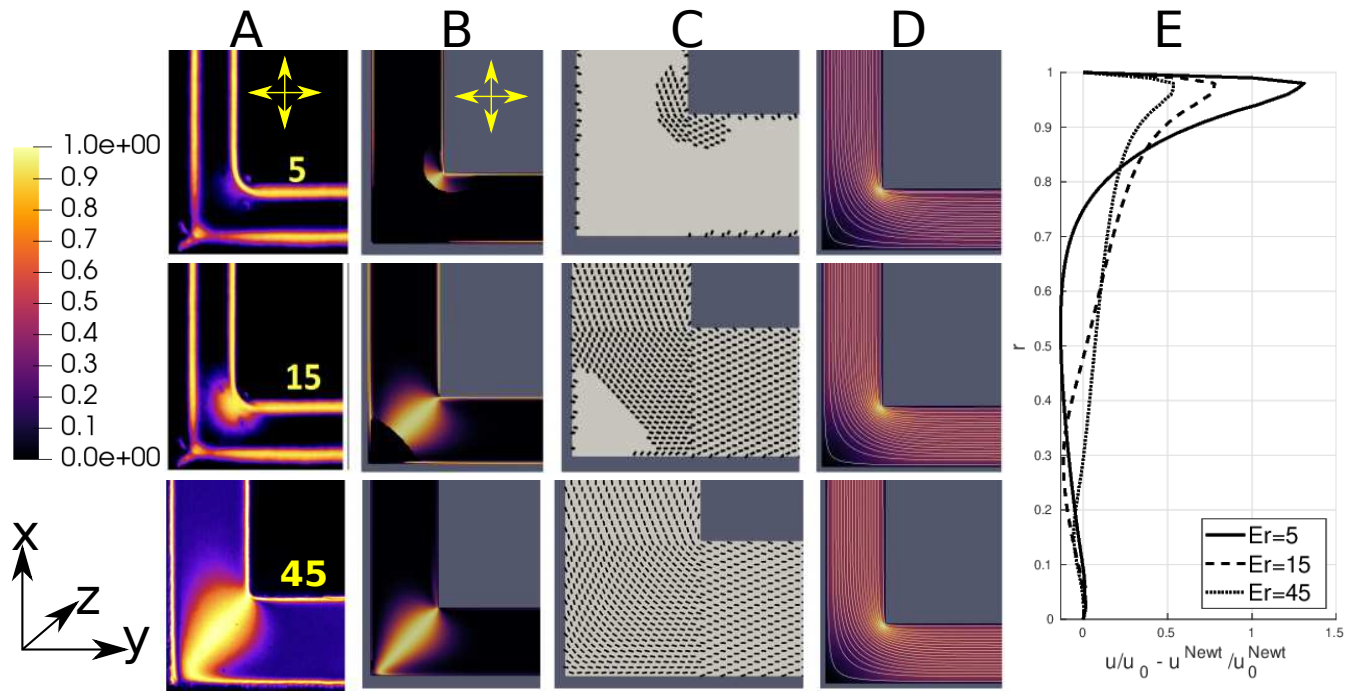}
\caption{\label{L_combined4} Director alignment in a flow through the L-channel. (A) Intensity image obtained with crossed polarizers. (B) Numerical prediction of intensity with crossed polarizers (calculated as $4n_x^2n_y^2$). (C) Glyphs of the director field. (D) Contours of flow magnitude scaled by the maximum flow speed. The Ericksen number is denoted by the white number in the lower left corner of column A pictures. (E) Effect of nematic liquid crystals on the amplification of flow flow asymmetry (in the symmetry of the U-bend) with respect to a Newtonian fluid. $r$ denotes the normalised distance from the outer corner.}
\end{figure*}

\subsection{ U-bend  }

The correlation between velocity gradients and flow-director alignment given by eq. (\ref{director_balance}) has been qualitatively confirmed by experimental measurements and numerical calculations. The director aligns in the $z-$direction at $Er<<1$, as indicated by the uniform dark signal throughout the U-bend (fig. \ref{paper_fig1_3}D). At $Er=O(1)$ the director orientation is also affected by viscous effects (eq. (\ref{director_balance})). In creeping flows, the peak velocity is displaced towards the bend axis \cite{fedorowicz2022, Fedorowicz2022_pof}. The resulting elevated shear rate at the inner wall is responsible for the emergence of the flow-aligned director field at the inside of the bend at $Er=4$. The result is experimentally manifested through a non-zero light intensity signal near the inner wall (upper row in fig. \ref{combined_u_bend2}A), and is qualitatively confirmed by numerical simulations (fig. \ref{combined_u_bend2}B, C). The size of the flow-aligned region scales with the Ericksen number, and by the time $Er=6$, the region of flow-director misalignment is confined to the outer bend wall (middle row in fig. \ref{combined_u_bend2}), wherein there is consequently relatively little flow. Further evidence for the shear-induced propagation of the flow-aligned region towards the outer bend side is provided by the light intensity plots across the bend that are given in the supplementary material.

The combination of nematic viscoelasticity and the geometric curvature produces variable director alignment across the channel. Due to viscous anisotropy, more flow is promoted in regions where the flow alignment is achieved. Fig. \ref{combined_u_bend2}D shows that the uneven flow distribution introduced by the geometry is amplified by the presence of nematic liquid crystals. That provides novel opportunities for flow partitioning, which can be controlled by Ericksen number. Flow tunning capabilities through viscous anisotropy are weaker at high Ericksen numbers, where the dominance of viscous effects results in complete flow alignment (lower row in fig. \ref{combined_u_bend2}) and thus nearly uniform effective viscosity. The corresponding velocity profile (dotted line in fig. \ref{combined_u_bend2}D) is much more similar to the Newtonian result than at intermediate $Er$.

\subsection{ L-channel   }

The physics governing microstructure arrangement in an L-channel is qualitatively similar to the U-bend. Here, the change in the flow direction occurs over a smaller length scale, so flow-aligning effects are more localised, as can be seen experimentally in fig. \ref{L_combined4}A and numerically in fig. \ref{L_combined4}B,C. Higher flow speeds occur near the inner corner, and the crossed polarizer images indicate that the director alignment is initiated there. A positive feedback mechanism is established by reducing the effective viscosity and promoting further flow in the inner corner (indicated by tighter streamline concentrations near the inner corner). At $Er=15$ the director remains unaligned with the flow only in the outer corner of the elbow, which results in a stagnation region there. The L-bend is thus seen to act like a microfluidic valve, where the effective valve throat is controlled by the Ericksen number. 

The effect of flow asymmetry is reduced at high Er, where flow dynamics dominate director alignment and the overall rheology tends towards Newtonian fluids. Our observation is confirmed in fig. \ref{L_combined4}D, which shows that the deviation of the velocity profile from Newtonian behaviour is smallest at high $Er$. Across a range of Ericksen numbers, flow rate control can be achieved by managing the size of the stagnation region, thus providing opportunities for novel flow-tuning concepts.

\subsection{History effects}

\begin{figure}
\includegraphics[width=1\linewidth]{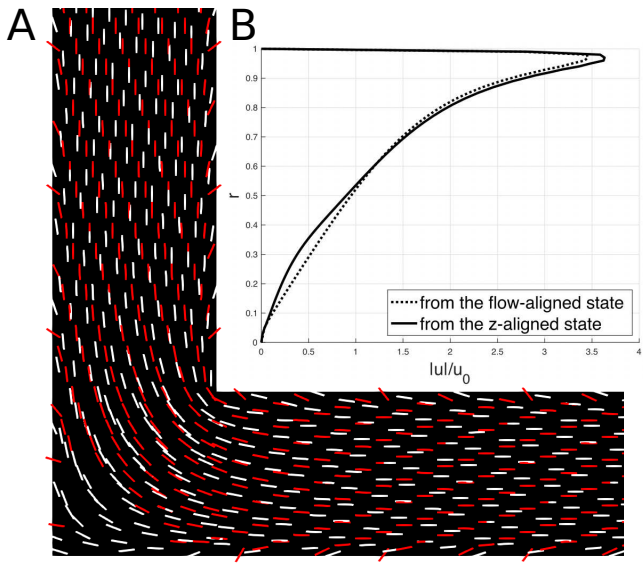}
\caption{\label{fig_hist_eff} (A) Comparison of director arrangements when the flow is initialised from the static (red) and flow aligned (white) states at $Er=15$. (B) Comparison of velocity profiles for differing initial conditions. }
\end{figure}

Figure \ref{fig_hist_eff}A demonstrates that NLCs carry history effects arising from previous material deformations, similar to other non-Newtonian fluids \cite{larson1999}. A flow-aligned state obtained by initialising the flow at higher $Er$ is preserved when the Ericksen number is subsequently reduced. This director arrangement differs from that obtained when the flow is initialised in the static ($z-$aligned) state (red rods in fig. \ref{fig_hist_eff}A). The historical evolution of the director produces different velocities local to both corners (\ref{fig_hist_eff}B). In particular, the size of the stagnation region is reduced when the director aligns with the flow throughout the domain. Since the director alignment is uniform, the anisotropic viscosity of liquid crystals cannot be exploited to control the size of stagnation region near the outer corner. The dependence of the final state on the deformation history is only relevant at intermediate Ericksen numbers; memory effects disappear at $Er<1$, where the director arrangement is governed solely by elasticity. 

\section{Summary}

We report that nematic liquid crystals flowing through curvilinear microchannels engender transverse gradient of the director field, which can feedback into the flow field, thus leading to a curvature-durector-flow coupling phenomenon. Under specific regimes of Ericksen numbers, this coupling can create abrupt flow changes, thereby inducing topological defects, and generate binary flow states. Flow alignment of the director field is promoted (hindered) in locations of higher (lower) flow speeds. The effect becomes stronger as the geometric curvature increases, and produces a positive feedback effect: as the director alignment increases, the apparent viscosity reduces, along with the reduction of the boundary layers. This increases the velocity gradients, thus further promoting the director alignment. The switching phenomenon can be potentially used to develop flow-controlled microfluidic valves. Finally, we found that the flow history controls the director arrangement at intermediate Ericksen numbers. A system that is in a fully aligned state initially, will tend to retain that state. On the contrary, initialising the flow from the static director equilibrium results only in a partial flow alignment. The finding could guide development of potential applications, for instance, non-intrusive flow actuation valves, LC-based microfluidic logic gates, and optofluidic switches and velocimetry devices. \\


\begin{acknowledgments}
\textbf{Acknowledgements} KF and RP acknowledge the EPSRC project the Centre in Advanced Fluid Engineering for Digital Manufacturing (Grant No. EP/R00482X/1) and Unilever plc for financial support. KF and RP would like to acknowledge the assistance given by Research IT and the use of the Computational Shared Facility at The University of Manchester. AS thanks the Luxembourg National Research Fund's ATTRACT Investigator Grant (Grant no. A17/MS/ 11572821/MBRACE) and CORE Grant (C19/MS/13719464/TOPOFLUME/Sengupta) for supporting this work. 

\end{acknowledgments}


\bibliography{bibFile2}

\end{document}